\documentclass[prl,aps,superscriptaddress,showpacs,preprint]{revtex4}
\usepackage{amsbsy}
\usepackage{amssymb}
\usepackage[dvips]{graphicx}

\begin{document}
\title{
Finite size scaling in
Villain's fully frustrated model
and singular effects of plaquette disorder}

\author{J. Lukic}
\affiliation{Dipartimento di Fisica, SMC and UdR1 of INFM,
Universit\`a di  Roma {\em La Sapienza}, P.le Aldo Moro 2, 00185 Roma, Italy.}

\author{E. Marinari}
\affiliation{Dipartimento di Fisica, INFN, 
Universit\`a di  Roma {\em La Sapienza}, P.le Aldo Moro 2, 00185 Roma, Italy.}

\author{O. C. Martin}
\affiliation{Laboratoire de Physique Th\'eorique et Mod\`eles Statistiques,
b\^atiment 100, Universit\'e Paris-Sud, F--91405 Orsay, France.}

\date{\today}

\begin{abstract}
The ground state and low $T$ behavior of two-dimensional spin systems
with discrete binary couplings are subtle but can be analyzed using
exact computations of finite volume partition functions.  We first
apply this approach to Villain's fully frustrated model, unveiling an
unexpected finite size scaling law. Then we show that the introduction
of even a small amount of disorder on the plaquettes dramatically
changes the scaling laws associated with the $T=0$ critical point.
\end{abstract}

\maketitle

Two-dimensional frustrated systems typically have a high ground-state
degeneracy: this can lead to unusual physical properties~\cite{Diep}.
In the absence of disorder, the zero-temperature spin states can 
often be mapped~\cite{Forgacs80} to 
Baxter type or solid on solid models, leading
to a relatively good understanding of the low temperature regime.
Unfortunately, the incorporation of
disorder renders analytic approaches powerless and one 
must resort to numerical treatment. In 
the limit of very strong disorder, one expects to have
a spin glass, at least at zero temperature. It has recently been shown
that the critical behavior of two-dimensional spin
glasses with binary quenched random couplings is rather 
subtle, being very different from the one suggested
from the na\"{\i}ve low temperature 
expansion~\cite{WangSwendsen88,SaulKardar93,noisg,joerg}.
In this work, we focus on the change of critical behavior
of the $2d$ fully frustrated model (FFM) when one adds
disorder.
In the FFM each elementary plaquette of
the lattice is frustrated, and
the couplings between the spins are all of the 
same magnitude. At $T=0$ (the model's critical point),
the FFM exhibits a power-law decay of spin-spin correlation
functions.
Furthermore, for $T>0$, the correlation length diverges exponentially
in $1/T$. Suppose now we allow some plaquettes to
be unfrustrated; does this reduction in the frustration
leave the critical properties invariant, or is it instead
a strong perturbation? Using exact partition functions
on finite lattices, we shall show that even small amounts
of disorder dramatically change the thermodynamic singularities
of the model.

\paragraph*{Villain's Fully Frustrated model ---} 
We consider a two dimensional square lattice with Ising spins
on the sites and binary couplings $J_{ij}$ on the bonds;
without loss of generality, we set $|J_{ij}|=J=1$. The Hamiltonian is
\begin{equation}
\label{eq:H}
H(\{\sigma_{i}\}) \equiv
  -\sum_{\langle ij \rangle} J_{ij}\;\sigma_{i}\;\sigma_{j}\;,
\end{equation}
where the sum runs over all pairs of nearest neighbor sites and
boundary conditions are periodic. We shall consider both $L \times L$
lattices and the strip geometry ($L \times \infty$).  In the FFM the
product of the four couplings on each elementary plaquette is $-1$:
there are many different ways to do this and most of them are gauge
equivalent.  More precisely, for a fully frustrated lattice with
periodic boundary conditions, there are four equivalence classes,
corresponding to whether the product of the $J$ along a loop winding
around a direction of the lattice is $+1$ or $-1$. Following
Villain~\cite{VILLAIN}, we choose the periodic implementation where
all $J$s are set to $1$ except on the vertical bonds where lines of
couplings alternate between $+1$ and $-1$ values.
Villain applied the transfer
matrix formalism to extract the free energy density in the infinite
volume limit:
\begin{eqnarray}
\label{eq:f_Villain}
& & -\beta f_{\infty}(\beta) =  \ln (2 \cosh (\beta J)) 
+ \frac{1}{16 \pi^{2}} \\
& & \int_{0}^{2 \pi} dh \int_{0}^{2 \pi} dk\; \ln[(1+z^2)^2
 - 2 z^2 (\cos 2h + \cos 2k)]\;. \nonumber  
\end{eqnarray}
Here $z=\tanh(\beta J)$, $\beta\equiv T^{-1}$ is the inverse
temperature, and $f_{\infty}$ is the free energy per site in the
infinite volume limit.

The free energy turns out to be analytic everywhere except at $T=0$.
The $T=0$ entropy density of the system (normalized to the number of
sites), $s_0$, is finite: $s_0=C/\pi$, where $C$ is the Catalan
number\cite{FISHER,WHO}.  The ground state energy density $e_0$ is
$-J$ (one quarter of the links are unsatisfied), and the low $T$
expansion of Eq.~(\ref{eq:f_Villain}) to leading order is
\begin{equation}
\label{eq:f_FFM_leading}
\beta f_{\infty}(\beta) \simeq
\beta e_0 - s_0 + c_1 \beta e^{-4 \beta J}\;.
\end{equation}
This makes clear the non-analyticity of the free energy at the
critical point $T=0$: a well behaved low temperature expansion
of $\beta f_{\infty}$ would give a series in $y=\exp(-4 \beta J)$ while
in expression (\ref{eq:f_FFM_leading}) 
a factor $\beta$ (i.e., a factor $\ln(y)$)
multiplies the leading contribution $y=e^{-4\beta J}$.

Taking the second derivative with respect to $T$
gives the leading behavior of the
specific heat density:
\begin{equation}
\label{eq:cv_FFM}
c_V(\beta) \simeq 16\; c_1\; J^2 \beta^3\,e^{-4 \beta J} \;.
\end{equation}
In the present case of the FFM, Eq.~(\ref{eq:f_Villain}) 
tells us everything about the
temperature scaling of thermodynamic quantities.
But it is also possible to obtain the free energy in 
a \emph{finite} volume: we shall use this to extract
both the correlation length and the finite size scaling.

\paragraph*{The FFM in a finite volume or on a strip ---}
Computing the partition function of the FFM in a finite volume is not
difficult.  For a finite lattice with periodic boundary conditions,
the partition function associated to the Hamiltonian of
Eq.~(\ref{eq:H}) can be expressed as the sum of four
Pfaffians~\cite{Zecchina1,Zecchina2,GalluccioLoebl00}, 
each multiplied by a plus or
minus sign: because of these signs, there are large
cancellations when combining the contributions of the four Pfaffians.
Thus it is necessary to compute each Pfaffian to very high precision;
we have done so with the ``Mathematica'' program that allows
for arbitrary precision computations. Basically, in this approach, the
double integral of Eq.~(\ref{eq:f_Villain}) becomes a sum over a
discrete set of momenta, where the support of this set is different
for the four Pfaffians. We have also
considered a strip geometry, where the vertical 
direction (where couplings are $\pm 1$) is infinite and the
other has a width $L$. In this case we have to compute mixed
sums and integrals from which we obtain 
the free-energy density of the FFM in
the strip geometry (typically we get $15$ or more significant digits).
We now focus on how the thermodynamic limit is
reached with increasing $L$.

\paragraph*{The correlation length ---}
The diverging correlation length of a physical system approaching
criticality can be defined in different ways: for example
one can use the exponential decay of spin-spin correlations or
the exponential convergence of the free energy density
with increasing lattice size. Since in our approach we do not have
access to the spin-spin correlation functions, we use the second
definition. Consider the strip geometry associated with a lattice of
infinite length in one direction and of width $L$ in the other (with
periodic boundary conditions): the correlation length $\xi$ can be
defined via the relation $(f_L - f_{\infty}) \sim e^{-L/\xi}$, where
the exponential can have a prefactor that depends smoothly on $L$, 
for example via a power law.

\begin{figure}
\centering
\includegraphics[width=5cm,height=7cm,angle=270] {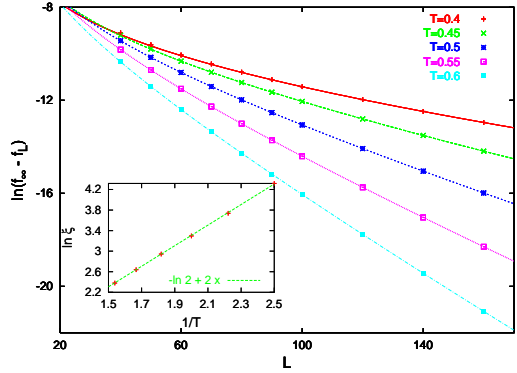}
\caption{ $\ln(f_{\infty}-f_L)$ as a function of $L$ (points)
and the best fits (continuous lines). In the inset,
$\ln(\xi)$ as a function of $1/T$.  
\protect\label{fig:FFM_bello}}
\end{figure}

In Fig.~(\ref{fig:FFM_bello}) we show 
$\ln(f_{\infty}-f_L)$ as a
function of $L$ (points) and the best fits to the form $a(T) -
m(T) L - c(T) \ln ( L )$ (continuous lines), for five
temperatures ($T\in [0.4, 0.6]$). 
$m(T)\equiv 1/\xi(T)$ is the inverse
correlation length.  The
quality of the fits is excellent, and we find that $c(T)$ 
depends only weakly on $T$ and is close to $1.5$.
Because of that, it is appropriate
to parametrize the $L$ dependence via the correlation length $\xi$
through the relation
\begin{equation}
\label{eq:diff_f_FFM}
f_L - f_{\infty} = A(T)\; \frac{e^{-L/\xi}}{L^{1.5}}\;,
\end{equation}
where $A(T)$ is a smooth function of $T$. From this we can 
extract $\xi$ when $L$ is large;
since our numerical values of $f_L - f_{\infty}$
loose precision beyond $L=200$, and getting higher precision would
have demanded a very large computational effort, we have been able to
extract $\xi$ by curve fitting only for $T \ge 0.4$.
In the inset of Fig.~(\ref{fig:FFM_bello}) we plot $\ln(\xi)$ as a
function of $1/T$, and find a straight line with slope very
close to $2$. Similarly, the prefactor of the exponential
is very close to $\frac12$ and we conjecture that this is in fact the
exact value (note that this prefactor has not been
estimated previously). We thus conclude that
\begin{equation}
\label{eq:xi_FFM}
\xi(T) \simeq e^{2 J \beta}/2\;.
\end{equation}
This scaling form applies as $T\to 0$, but it holds to
good accuracy even when $\xi$ is not so large: 
for example at $T=1$ where $\xi\approx 4$ the value given by
(\ref{eq:xi_FFM}) is only $10\%$ off from the actual measured $\xi$.
In the standard hyperscaling framework, the singular part of $\beta
f_{\infty}$ is given by $\xi^{-d}$ for a d-dimensional model (up to
constants and possible logarithmic terms in $\xi$).  Using
Eq.~(\ref{eq:xi_FFM}), we then expect the singular part of
$f_{\infty}$ to go as $\exp(-4 \beta J)$, which is what was found from
Eq.~(\ref{eq:f_FFM_leading}): in the FFM, hyperscaling holds.

\paragraph*{Finite size scaling ---}
We have discussed the $1 \ll \xi \ll L$ region. It is also 
of interest to consider the 
regime where $L \ll \xi$: only by controlling this region
we can reach a full understanding of the finite size scaling of the
system.

As a start, we notice that good data collapse is obtained as $T \to 0$
when we multiply $(f_L - f_{\infty})$ by the factor
$\beta L^2$, i.e., we find
\begin{equation}
\label{eq:df_FFM_FSS}
\beta L^2 (f_L - f_{\infty}) \simeq W(L/\xi) \ ,
\end{equation}
where $W(L/\xi)$ is an adimensional function of the ratio $L/\xi$.
When $x=L/\xi\to \infty$ we recover the previous analysis and $W(x)
\approx \sqrt{x} \exp(-x)$.  In the opposite limit, $x=L/\xi\to 0$,
the function $W$ goes to a constant ~\footnote{This is connected to the
fact that finite volume corrections to the ground state entropy are of
order $L^{-2}$. In $(f_L-f_\infty)$, the term $e_0$ has no
corrections, while $s_0(L)\sim s_0(\infty) + \delta s_0 / L^2$, and
$\delta s_0$ is the relevant constant.}.
Using the relations
(\ref{eq:f_FFM_leading}),
(\ref{eq:diff_f_FFM}) and 
(\ref{eq:df_FFM_FSS}), we find
$$
f_L \simeq e_0 -\frac{s_0}{\beta}+c_1e^{-4\beta J}+\frac{W}{\beta L^2}\;.
$$
Dividing and multiplying the last term by $\xi^2$ and
substituting $2\beta$ by $\ln(\xi)$ (cf. Eq.~(\ref{eq:xi_FFM}))
we obtain
\begin{equation}
\label{eq:final_scaling}
f_L \simeq e_0 -\frac{s_0}{\beta}+c_1 e^{-4\beta J}
\left(
1+\widetilde{W}\left(\frac{L}{\xi}\right)
/\ln(\xi)
\right)
\end{equation}
where $\widetilde{W}$ is a scaling function simply related to $W$.
The important point is that
in Eq. (\ref{eq:final_scaling})
the adimensional scaling function $\widetilde{W}(L/\xi)$ 
is divided by a factor $\beta\sim\ln(\xi)$, so one has 
an \emph{anomalous} finite size scaling law.

\begin{figure}
\includegraphics[width=5cm, height=8cm,angle=270]{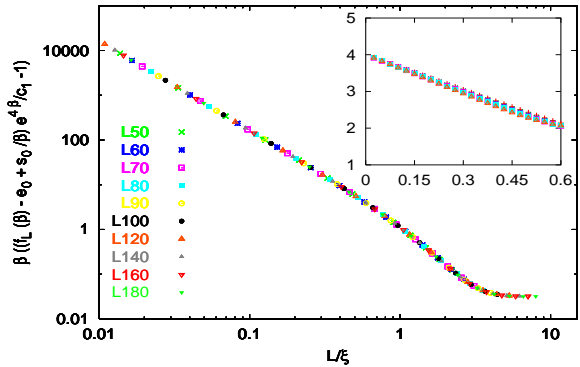}
\caption{
$ \beta ( (f_L(\beta)-e_0+\frac{s_0}{\beta})
e^{4\beta J} / c_1 - 1)$ 
versus $L/\xi$.
Here $c_1=-1.273$. Inset: $-T \ln(T^2 c_V)$ vs. $T$
showing the convergence to $A=4$ (cf. section with disorder).
\protect\label{fig:FFM_FSS}}
\end{figure}

The function $\widetilde{W}$ satisfies
\begin{displaymath}
\widetilde{W}(x)=\left \{ \begin{array}{ll}
x^{-2} & \mbox{when } x \to 0 \;,\\
\mbox{constant } & \mbox{when } x \to \infty\;.
\end{array} \right.
\end{displaymath}
We show all of our data for 
$ \beta ( (f_L(\beta)-e_0+\frac{s_0}{\beta})
e^{4\beta J} / c_1 - 1)$ 
versus $L/\xi$ in Fig.~(\ref{fig:FFM_FSS}).
Here
$10 \le L \le 180$ and 
$0 < T \le 1.0$. 
$c_1$ has been determined with a best fit to $f_\infty$ and has the
value
$c_1=-1.273$.
The data collapse is excellent and since it involves
a \emph{correction} to $f_L$, one can conclude that
finite size effects are under very good control.

To complete our study of the FFM without disorder, 
we consider finally the low temperature expansion of $f_L(\beta)$:
\begin{equation}
f_L(\beta) \simeq e_0 -\frac{s_0}{\beta}
-\frac{1}{\beta L^2} \frac{g_1}{g_0}\exp(-4\beta J)\;,
\label{eq:low-t-exp}
\end{equation}
where $g_0$ and $g_1$ are the respectively the degeneracy of the
ground state and of the first excited state. We determine $g_0$ and
$g_1$, finding that to very good accuracy
\begin{equation}
g_1/g_0 \simeq A L^2 + B L^2\ln(L)\;,
\label{eq:g-scale}
\end{equation}
with $A=-0.44$ and $B=0.63$. It is possible to show that the scalings
(\ref{eq:g-scale}) and (\ref{eq:final_scaling}) are mutually
compatible. 

\paragraph*{Adding disorder / diluting frustration in the FFM ---} 
At this point the large $L$ and $\xi$ behavior of the FFM is well
understood. Now we move on to see what happens when frustration is
partly removed. We do this by unfrustrating a small fraction of the
plaquettes, choosing these at random. The set of couplings $J_{ij}$
entering the Hamiltonian (\ref{eq:H}) is now such that on a fraction
$p_1$ of the plaquettes their product is equal to $1$.  We shall refer
refer to this as the Plaquette Disorder (PD) ensemble.  Does this
modification to the Hamiltonian change the scaling laws of the FFM?
Interestingly, the change is in fact dramatic, as we now reveal.

\paragraph*{The computational tool ---}
The presence of plaquette quenched disorder breaks the translation
invariance of the Hamiltonian; because of this, the transfer matrix
cannot be diagonalized by going to Fourier space. Instead, we rely on
the explicit computation of Pfaffians; this can be done for any set of
$J_{ij}$.  A very effective approach for computing such Pfaffians,
based on {\em modular arithmetic}, has been proposed and implemented
in~\cite{GalluccioLoebl00}. One evaluates the partition function in a
(large) finite volume via its low temperature series:
\begin{equation}
\label{eq:Zbeta}
Z_J\left(\beta\right)=e^{2 L^2 \beta J} 
\;P_J\left(e^{-2\beta J}\right)\;,
\end{equation}
where $P_J(x)$ is a polynomial whose integer coefficients are the
number of spin configurations of a given energy.  The
algorithm~\cite{GalluccioLoebl00} determines these integers
\emph{exactly}, allowing one to analyze the system even for very low
temperatures.

In our implementation \cite{GalluccioLoebl00,noisg}
the CPU time to compute $Z_J$ grows approximately as
$L^{5.5}$. We have mainly studied the model with an unfrustrated
fraction of plaquettes $p_1=1/8$, by determining $Z_J$ on lattices
with sizes ranging from $L=24$ ($2000$ samples) up to $L=56$ 
($200$ samples). We have also analyzed
the case of $p_1=1/4$ on lattices with $L=32$ ($600$
samples) up to $L=48$ ($100$ samples).  We computed 
sample averages of different physical quantities like the
free energy and the specific heat, and we analyzed $T=0$
quantities like the number of ground states and of low-lying excited
states.

\paragraph*{Low temperature scaling of $c_V$ ---}
In the following we mainly analyze the specific heat, $c_V$, that we
compute from the fluctuations of the internal energy~\cite{noisg}.

In the case without disorder, the low $T$ thermodynamic limit
behavior (where $V$ diverges at fixed low $T$) of $c_V$ is given
in Eq.~(\ref{eq:cv_FFM}). On the contrary in the finite size limited,
$T\to 0$ limit (where $T\to 0$ at fixed volume), we have that
\begin{equation}
\label{eq:cv_scaling_naive}
c_V \equiv 
\frac{\beta^2}{L^2}
\left\langle \  \left[ H - \langle H \rangle \right]^2\right\rangle \;
\thickapprox
\frac{16\,\beta^2J^2}{L^2}
\frac{g_1}{g_0}
e^{- 4 \beta J}\;.
\end{equation}
Thus, the scaling
in the thermodynamic limit
differs from the finite size limited, 
finite volume low $T$ scaling, but only in
the power of the $\beta$ prefactor, and not in the argument of the
exponential. (This power 
is related to the logarithmic
corrections in the $L$ scaling of the ratio $g_1/g_0$ of Eq.
(\ref{eq:g-scale})). Disorder completely
changes this picture as we shall now make clear.

\begin{figure}
\includegraphics[width=5cm, height=7cm,angle=270]{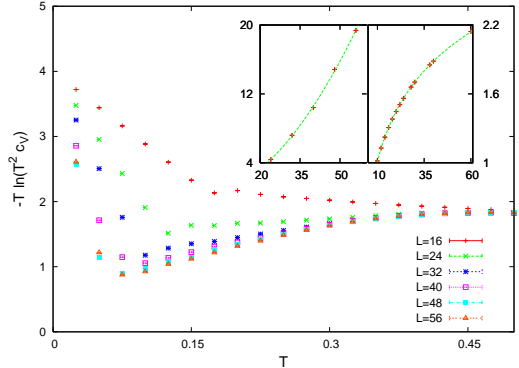}
\caption{$-T \ln(T^2 c_{V})$ versus $T$ in the PD model
with a fraction $1/8$ of unfrustrated plaquettes.
In the inset:
on the left,
$\ln({g_1}/{g_0 L^2})$ versus $L$ for the PD model;
on the right,
${g_{1}}/{g_{0} L^2}$ versus $L$ for the FFM.
\protect\label{fig:DIL_PD}}
\end{figure}

Let us parametrize the scaling behavior by the argument of the exponential
function and by a power for the first subleading correction:
$c_V\thickapprox\beta^P\;e^{-A\beta J}$.
We have seen in the FFM that $A=4$ and that $P=2$ in the na\"{\i}ve,
finite size limited scaling while $P=3$ in the scaling regime
of the thermodynamic limit.  In
Fig.~(\ref{fig:DIL_PD}) we plot $-T \ln\left(T^Pc_V\right)$ versus $T$
for the PD model (with $p_1=1/8$).  We take here $P=2$, and try to
determine $A$.  We proceed in this way since the value of $A$ is not
very sensitive to the value that we assume for $P$: our conclusions
for the value of $A$ would be unchanged by taking $P$ up to
$10$, and would be strengthened by taking $P<2$.  $A$ is given by the
intercept of the envelope's extrapolation to $T=0$. We can distinguish
three regions in Fig.~(\ref{fig:DIL_PD}). The region at very
low $T$ values corresponds to the na\"{\i}ve scaling with $A=4$ 
as can be seen from the $T=0$ intercept; this
region shrinks to zero with
increasing lattice size. There is also the high $T$ region but
it is not relevant
for critical properties. Finally, 
the most interesting region is in between the
other two; there, one has the scaling \emph{in the thermodynamic limit},
obtained
from the envelope of the set of curves. It is clear
from the figure that this scaling is incompatible with
$A=4$; in fact, the estimated value for $A$ decreases 
with the lattice size
(since for higher $L$ we can determine the envelope to lower
$T$ values), and possibly goes to zero when $L\to\infty$: such
a behavior 
would imply an algebraic $T \to 0$ limit, and not an exponential
singularity (it is clear that if this happens $P$ will become smaller
than zero: our best fits are already compatible with a small negative
value of $P$, but since they are
not very sensitive to the value of $P$ we
cannot give a quantitative estimate of this effect). 
We have repeated this
analysis for a dilution of $1/4$ with similar conclusions: again the
curves go down to low values of $A$, possibly $0$.
As a final note and to drive home even more the incredibly
strong effects of disorder, one can compare the
behavior of $c_V$ in the pure and the PD models:
as shown in the inset of Fig.~(\ref{fig:FFM_FSS})
the pure model has both a very clear exponential low temperature limit
and small finite size effects; on the contrary in the presence
of disorder, $c_V$ has large finite size effects
and the $\exp(-4 \beta J)$ scaling is clearly absent.

\paragraph*{The degeneracy of low energy states ---}
In the insets of Fig.~(\ref{fig:DIL_PD}) we plot ${g_{1}}/{g_{0}L^2}$ 
as a function of $L$ for the FFM and $\ln({g_1}/{g_0L^2})$ 
(that we get as an output of our
exact computation of $Z$)
as a function of $L$ for the PD: this is because the first
ratio grows logarithmically in $L$ (see Eq.~(\ref{eq:g-scale})),
while in the presence of quenched disorder, it seems to grow
at least exponentially fast with $L$. 
The plaquette disorder completely changes the scaling of these
quantities and so the
low temperature expansion for the pure and the 
disordered models will be very different.

\paragraph*{Summary and discussion ---}
Two dimensional frustrated models, with or without disorder, are
challenging systems, especially when they have a critical point as
Villain's fully frustrated model does. We first studied in depth his
(pure) model: beyond a conjecture for the magnitude of the correlation
length, we showed that finite size scaling was \emph{anomalous}. We
also found that the ratio $g_1/g_0$ of the degeneracies of the lowest
energy states grows as $L^2$ with \emph{multiplicative} logarithmic
corrections: as a result, the power law in $T$ multiplying the
$\exp(-4\beta J)$ scaling of the specific heat $c_V$ is modified.

We then introduced quenched disorder in the form of a small fraction
of randomly positioned unfrustrated plaquettes to find that
qualitatively new phenomena arise. For instance, $g_1/g_0$ grows
exponentially in $L$ rather than as a power. Following the argument of
what occurs in the pure model, this growth breaks the $\exp(-4\beta J)$
scaling of $c_V$, taking one to a form of the type $\exp(-A\beta J)$
with $A$ rather small if non-zero.  These effects are striking and
show the extreme fragility of the pure system: the universality class
of the FFM is completely changed when disorder is introduced.

We thank T. J\"org for stimulating discussions. The
code for computing the partition function is based on the original
code by A. Galluccio, M. L\"obl, G. Rinaldi and J. Vondr\'ak. Work
supported by the EEC's FP6 Information Society Technologies Programme
under contract IST-001935, EVERGROW (www.evergrow.org), and by the
EEC's HPP under contracts HPRN-CT-2002-00307 (DYGLAGEMEM) and
HPRN-CT-2002-00319 (STIPCO).  The LPTMS is an Unit\'e de Recherche de
l'Universit\'e Paris~XI associ\'ee au CNRS.

\bibliographystyle{apsrev}
\bibliography{ref}

\addcontentsline{toc}{chapter}{\protect\bibname}
\begin{thebibliography}{12}
\expandafter\ifx\csname natexlab\endcsname\relax\def\natexlab#1{#1}\fi
\expandafter\ifx\csname bibnamefont\endcsname\relax
  \def\bibnamefont#1{#1}\fi
\expandafter\ifx\csname bibfnamefont\endcsname\relax
  \def\bibfnamefont#1{#1}\fi
\expandafter\ifx\csname citenamefont\endcsname\relax
  \def\citenamefont#1{#1}\fi
\expandafter\ifx\csname url\endcsname\relax
  \def\url#1{\texttt{#1}}\fi
\expandafter\ifx\csname urlprefix\endcsname\relax\def\urlprefix{URL }\fi
\providecommand{\bibinfo}[2]{#2}
\providecommand{\eprint}[2][]{\url{#2}}

\bibitem[{\citenamefont{Diep}(2005)}]{Diep}
\bibinfo{editor}{\bibfnamefont{H.~T.} \bibnamefont{Diep}}, ed.,
  \emph{\bibinfo{title}{Frustrated Spin Systems}} (\bibinfo{publisher}{World
  Scientific}, \bibinfo{address}{Singapur}, \bibinfo{year}{2005}).

\bibitem[{\citenamefont{Forgacs}(1980)}]{Forgacs80}
\bibinfo{author}{\bibfnamefont{G.}~\bibnamefont{Forgacs}},
  \bibinfo{journal}{Phys. Rev. B} \textbf{\bibinfo{volume}{22}},
  \bibinfo{pages}{4473} (\bibinfo{year}{1980}).

\bibitem[{\citenamefont{Wang and Swendsen}(1988)}]{WangSwendsen88}
\bibinfo{author}{\bibfnamefont{J.-S.} \bibnamefont{Wang}} \bibnamefont{and}
  \bibinfo{author}{\bibfnamefont{R.~H.} \bibnamefont{Swendsen}},
  \bibinfo{journal}{Phys. Rev. B} \textbf{\bibinfo{volume}{38}},
  \bibinfo{pages}{4840} (\bibinfo{year}{1988}).

\bibitem[{\citenamefont{Saul and Kardar}(1993)}]{SaulKardar93}
\bibinfo{author}{\bibfnamefont{L.}~\bibnamefont{Saul}} \bibnamefont{and}
  \bibinfo{author}{\bibfnamefont{M.}~\bibnamefont{Kardar}},
  \bibinfo{journal}{Phys. Rev. E} \textbf{\bibinfo{volume}{48}},
  \bibinfo{pages}{R3221} (\bibinfo{year}{1993}).

\bibitem[{\citenamefont{Galluccio et~al.}(2004)\citenamefont{Galluccio, Lukic,
  Marinari, Martin, and Rinaldi}}]{noisg}
\bibinfo{author}{\bibfnamefont{A.}~\bibnamefont{Galluccio}},
  \bibinfo{author}{\bibfnamefont{J.}~\bibnamefont{Lukic}},
  \bibinfo{author}{\bibfnamefont{E.}~\bibnamefont{Marinari}},
  \bibinfo{author}{\bibfnamefont{O.~C.} \bibnamefont{Martin}},
  \bibnamefont{and} \bibinfo{author}{\bibfnamefont{G.}~\bibnamefont{Rinaldi}},
  \bibinfo{journal}{Phys. Rev. Lett.} \textbf{\bibinfo{volume}{92}},
  \bibinfo{pages}{117202} (\bibinfo{year}{2004}).

\bibitem[{\citenamefont{J{\"o}rg et~al.}()\citenamefont{J{\"o}rg, Lukic,
  Marinari, and Martin}}]{joerg}
\bibinfo{author}{\bibfnamefont{T.}~\bibnamefont{J{\"o}rg}},
  \bibinfo{author}{\bibfnamefont{J.}~\bibnamefont{Lukic}},
  \bibinfo{author}{\bibfnamefont{E.}~\bibnamefont{Marinari}}, \bibnamefont{and}
  \bibinfo{author}{\bibfnamefont{O.~C.} \bibnamefont{Martin}},
  \bibinfo{note}{to be published}.

\bibitem[{\citenamefont{Villain}(1977)}]{VILLAIN}
\bibinfo{author}{\bibfnamefont{J.}~\bibnamefont{Villain}}, \bibinfo{journal}{J.
  Phys. C: Solid State Phys.} \textbf{\bibinfo{volume}{10}},
  \bibinfo{pages}{1717} (\bibinfo{year}{1977}).

\bibitem[{\citenamefont{Fisher}(1961)}]{FISHER}
\bibinfo{author}{\bibfnamefont{M.~E.} \bibnamefont{Fisher}},
  \bibinfo{journal}{Phys. Rev.} \textbf{\bibinfo{volume}{124}},
  \bibinfo{pages}{1664} (\bibinfo{year}{1961}).

\bibitem[{\citenamefont{Andr\'e et~al.}(1979)\citenamefont{Andr\'e, Bidaux,
  Carton, Conte, and de~Seze}}]{WHO}
\bibinfo{author}{\bibfnamefont{G.}~\bibnamefont{Andr\'e}},
  \bibinfo{author}{\bibfnamefont{R.}~\bibnamefont{Bidaux}},
  \bibinfo{author}{\bibfnamefont{J.-P.} \bibnamefont{Carton}},
  \bibinfo{author}{\bibfnamefont{R.}~\bibnamefont{Conte}}, \bibnamefont{and}
  \bibinfo{author}{\bibfnamefont{L.}~\bibnamefont{de~Seze}},
  \bibinfo{journal}{J. Appl. Phys.} \textbf{\bibinfo{volume}{50}},
  \bibinfo{pages}{7345} (\bibinfo{year}{1979}).

\bibitem[{\citenamefont{Regge and Zecchina}(1996)}]{Zecchina1}
\bibinfo{author}{\bibfnamefont{T.}~\bibnamefont{Regge}} \bibnamefont{and}
  \bibinfo{author}{\bibfnamefont{R.}~\bibnamefont{Zecchina}},
  \bibinfo{journal}{J. Math. Phys.} \textbf{\bibinfo{volume}{37}},
  \bibinfo{pages}{2796} (\bibinfo{year}{1996}).

\bibitem[{\citenamefont{Regge and Zecchina}(2000)}]{Zecchina2}
\bibinfo{author}{\bibfnamefont{T.}~\bibnamefont{Regge}} \bibnamefont{and}
  \bibinfo{author}{\bibfnamefont{R.}~\bibnamefont{Zecchina}},
  \bibinfo{journal}{J. Phys. A} \textbf{\bibinfo{volume}{33}},
  \bibinfo{pages}{741} (\bibinfo{year}{2000}).

\bibitem[{\citenamefont{Galluccio et~al.}(2000)\citenamefont{Galluccio, Loebl,
  and Vondr\'ak}}]{GalluccioLoebl00}
\bibinfo{author}{\bibfnamefont{A.}~\bibnamefont{Galluccio}},
  \bibinfo{author}{\bibfnamefont{M.}~\bibnamefont{Loebl}}, \bibnamefont{and}
  \bibinfo{author}{\bibfnamefont{J.}~\bibnamefont{Vondr\'ak}},
  \bibinfo{journal}{Phys. Rev. Lett.} \textbf{\bibinfo{volume}{84}},
  \bibinfo{pages}{5924} (\bibinfo{year}{2000}).

\end{thebibliography}

\end{document}